\documentclass[prl,twocolumn,superscriptaddress,showpacs,floatfix,longbibliography]{revtex4-1}
\usepackage{mathrsfs,braket}
\usepackage{amssymb, amsbsy, amsmath, latexsym, dsfont, array, layout,
graphicx,mathrsfs,color,ulem,bm}
\usepackage{cancel}
\usepackage[colorlinks=true,citecolor=blue,urlcolor=blue]{hyperref}

\begin{document}

\title{Simulating Floquet non-Abelian topological insulator with photonic quantum walks}

\author{Quan Lin}\thanks{These authors contributed equally to this work.}
\affiliation{Key Laboratory of Quantum Materials and Devices of Ministry of Education, School of Physics, Southeast University, Nanjing 211189, China}
\author{Tianyu Li}\thanks{These authors contributed equally to this work.}
\affiliation {Key Laboratory of Atomic and Subatomic Structure and Quantum Control (Ministry of Education), Guangdong Basic Research Center of Excellence for Structure and Fundamental Interactions of Matter, School of Physics, South China Normal University, Guangzhou 510006, China}
\affiliation {Guangdong Provincial Key Laboratory of Quantum Engineering and Quantum Materials, Guangdong-Hong Kong Joint Laboratory of Quantum Matter, Frontier Research Institute for Physics, South China Normal University, 510006 Guangzhou, China}
\author{Haiping Hu}
\affiliation{Beijing National Laboratory for Condensed Matter Physics, Institute of Physics, Chinese Academy of Sciences, 100190 Beijing, China}
\affiliation{School of Physical Sciences, University of Chinese Academy of Sciences, 100049 Beijing, China}
\author{Wei Yi}\email{wyiz@ustc.edu.cn}
\affiliation{CAS Key Laboratory of Quantum Information, University of Science and Technology of China, Hefei 230026, China}
\affiliation{CAS Center For Excellence in Quantum Information and Quantum Physics, Hefei 230026, China}
\author{Peng Xue}\email{gnep.eux@gmail.com}
\affiliation{Key Laboratory of Quantum Materials and Devices of Ministry of Education, School of Physics, Southeast University, Nanjing 211189, China}

\begin{abstract}
\bf{Floquet non-Abelian topological phases emerge in periodically driven systems and exhibit properties that are absent in their Abelian or static counterparts. Dubbed the Floquet non-Abelian topological insulators (FNATIs), they are characterized by non-Abelian topological charges and feature multifold bulk-boundary correspondence, making their experimental observation challenging. Here we simulate the FNATI using a higher-dimensional photonic quantum walk and develop dynamic measurement schemes to demonstrate key signatures of the FNATI. Importantly, combining a direct bulk-dynamic detection for the underlying quaternion topological charge, and a spatially-resolved injection spectroscopy for the edge states, we experimentally establish the multifold bulk-boundary correspondence, and, in particular, identify the anomalous non-Abelian phase where edge states appear in all band gaps, despite the presence of a trivial topological charge. Our experiment marks the first experimental characterization of the FNATI, providing general insight into the non-Abelian topological phases.}
\end{abstract}

\maketitle

Topological phases represent a novel state of matter that transcends the Ginzburg-Landau symmetry-breaking framework~\cite{qi,kane}. Classified by their underlying symmetries and spatial dimensions, gapped Hamiltonians are characterized by integer topological invariants that dictate the existence of topologically protected edge states, as outlined by the principle of the bulk-boundary correspondence~\cite{qi,kane,AZ,ryu,class1,class2,class3,hatsu,graf,wangz,zyang,longhi,rudms}.
The recent discovery of non-Abelian topological phases significantly enriches the scene, where the topology of multigap systems requires characterization through non-Abelian topological charges~\cite{wu,hqiu,guo,jiangtianshu,zhengz,bzd2020,jiang1,jiang2,yangyihao,slager2024,lihu,yangy,chengd}. While the non-Abelian topological phase has been experimentally demonstrated using transmission line networks~\cite{guo,jiangtianshu}, the underlying multigap topology, along with the non-Abelian description, have found broad applications in the topological defects of nematic liquids~\cite{nematic1,nematic2,nematic3,nematic4}, and non-Hermitian  multiband models~\cite{guocx,nhknot,nhbraexp,braid2}.

Novel non-Abelian topology also arises in periodically driven systems, exemplified by the recently proposed Floquet non-Abelian topological insulator (FNATI)~\cite{lihu}. Unlike its Abelian or static counterparts, for the FNATI, the non-Abelian topological charge does not suffice to fully characterize the information of edge states, since the bulk-boundary correspondence therein is multifold, with multiple distinct edge-state configurations corresponding to the same non-Abelian topological charge. Such a multifold bulk-boundary correspondence, together with the intrinsic complexity of the non-Abelian topological charge, makes the experimental demonstration of FNATI challenging.

In this work, we propose and experimentally implement a three-band discrete-time quantum walk governed by a Floquet operator, whose underlying topology is characterized by the quaternion group $Q_8$. For the experiment, photons are used as quantum walkers~\cite{xl1,lq1,lq2,sch}, where the higher-dimensional coin states are encoded in the combined polarization and spatial degrees of freedom of the photons. We then construct the quaternion topological charge from the bulk dynamics by performing state tomography of the walker, which yields information of the Floquet operator in the quasimomentum space. For the detection of topological edge states, we introduce a domain-wall configuration and develop a novel spatially resolved injection spectroscopy to accurately determine the spectral location of the edge states. These dynamic measurements enable us to establish the multifold bulk-boundary correspondence in our domain-wall configuration. We further demonstrate the presence of an anomalous non-Abelian topological phase using the quantum-walk dynamics, where edge states exist even when the topological charge of the quantum-walk system is trivial. This sharply contrasts with static systems where edge states cannot persist with a trivial topological charge. While our work is the first experimental characterization of the FNATI, the newly developed spatially-resolved injection spectroscopy offers a universal approach for studying complex topological phases using quantum-walk dynamics.

{\bf Results}

{\bf Quantum walks with non-Abelian topology.}
We implement a one-dimensional discrete-time quantum walk governed by the following Floquet operator
\begin{align}
U = U_h RSR U_h,\label{eq:U}
\end{align}
where $S= \sum_x |x\rangle\langle x-1|\otimes |A\rangle\langle A|+|x\rangle\langle x|\otimes |B\rangle\langle B|+|x+1\rangle\langle x|\otimes |C\rangle\langle C|$ is the shift operator with the coin states $|A\rangle$, $|B\rangle$, and $|C\rangle$ and the walker's position state $|x\rangle$, $U_h =\sum_x|x\rangle\langle x|\otimes e^{-iM/2}$ with $M =
\begin{pmatrix}
a& r&u\\
r& h&s\\
u&s&g
\end{pmatrix}$, and $R= \sum_x|x\rangle\langle x|\otimes
\begin{pmatrix}
\cos\theta& i\sin\theta&0\\
i\sin\theta& \cos\theta&0\\
0&0&e^{i\phi}
\end{pmatrix}$ are two rotation operators. The $3\times3$ matrices are in the basis $\{|A\rangle,|B\rangle,|C\rangle\}$, and all the parameters $r$, $u$, $s$, $a$, $h$, $g$, $\theta$ and $\phi$ are real numbers.

\begin{figure*}[th]
\includegraphics[width=\textwidth]{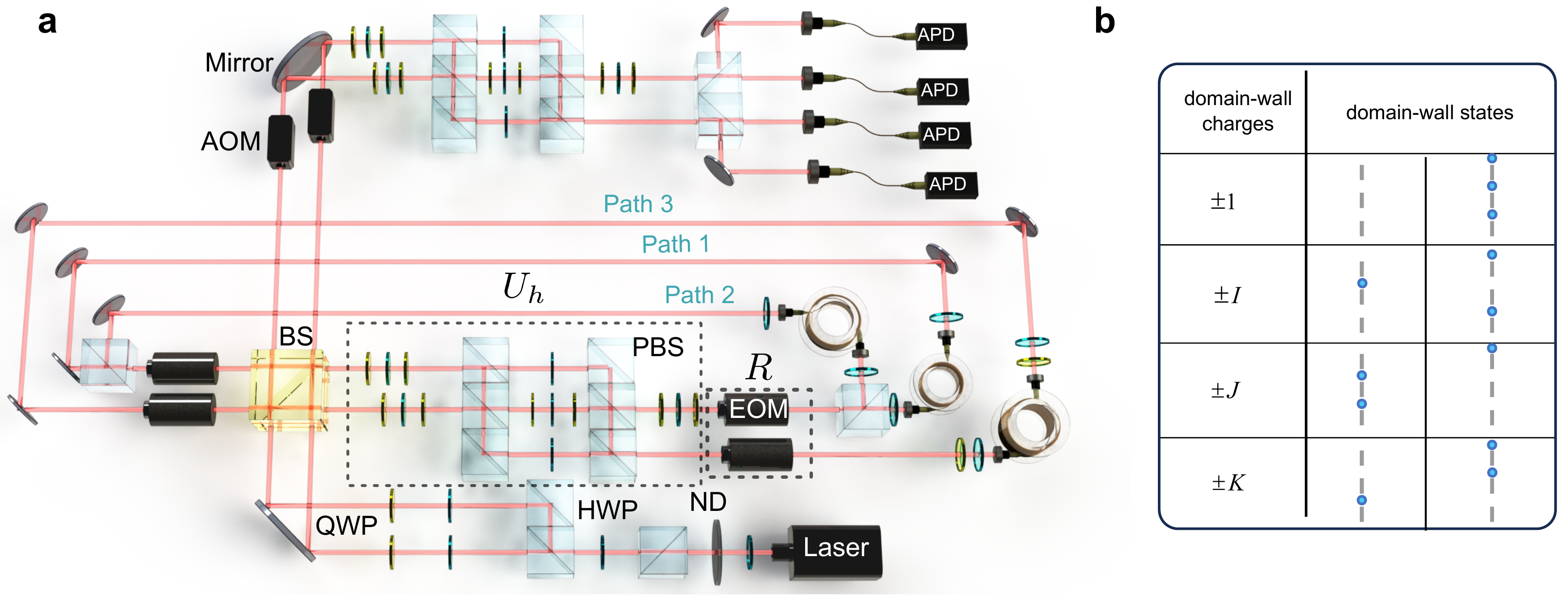}
\caption{{\bf Experimental setup and multiple bulk-boundary correspondence.} ({\bf A}) Experimental setup for observing the Floquet non-Abelian topological insulator via a three-band time-multiplexed photonic quantum walk. The photon source consists of a pulsed laser with a central wavelength of $808$nm, a pulse width of $88$ps, and a repetition rate of $15.625$KHz. Laser pulses are attenuated to the single-photon level through the use of a neutral density filter, resulting in an estimated average photon number per pulse of less than $3.8\times 10^{-4}$, which ensures a negligible probability of multiphoton events. Attenuated pulses are then coupled in and out of an interferometric network through a beam splitter (BS) with reflectivity of $3\%$. The rotation operator $U_h$ is realized by two triple polarizing beam splitters (PBSs) and waveplates. The rotation operator $R$  is carried out with two electro-optic modulators (EOMs) with different optical axis angles. The shift operator $S$  is realized by splitting the laser pulses into three single-mode fibers of varying lengths using two PBSs. In this setup, spatial modes are encoded into temporal shifts within a designated time step. The out-coupled photons are detected using avalanche photodiodes (APDs) in a time-resolved manner. An acousto-optic modulator (AOM) serves as an optical switch to protect the APDs by allowing photons to reach them only at the time of measurement. ({\bf B}) Multiple bulk-boundary correspondence. The first column lists the domain-wall charges, with the four classes of $Q_8$ represented in different rows. The second and third columns illustrate the patterns of the domain-wall states corresponding to the domain-wall charges specified in the first column. Blue dots indicate the presence of domain-wall states.}\label{fig1}
\end{figure*}

In the experimental implementation, we utilize a time-multiplexed scheme as shown in Fig.~\ref{fig1}A. Pulses from the photon source are attenuated to the single-photon level through the use of a neutral density filter, which ensures a negligible probability of multiphoton events. Necessitated by the multiband (multigap) condition of the model, the basis states of the three-dimensional coin are encoded into the hybrid polarization-spatial modes of photons, i.e., $\left\{|A\rangle=|UH\rangle, |B\rangle=|UV\rangle, |C\rangle=|DH\rangle\right\}$, where $|UH\rangle$ and $|UV\rangle$ represent horizontally and vertically polarized photons in the upper spatial mode, respectively, and $|DH\rangle$ represents vertically polarized photons in the lower spatial mode. Coin state preparation is realized by an interferometer with two triple polarizing beam splitters (PBSs)~\cite{methods}.

In the momentum space, the Floquet operator $U$ is transformed to $U_k$, which possesses space-time inversion ($\mathcal{PT}$) symmetry. The combination of the space-inversion $\mathcal{P}$ and time-reversal $\mathcal{T}$ symmetries can be represented by the complex conjugation operator $\mathcal{K}$ in an appropriate basis. That is, $\mathcal{K} U_k \mathcal{K} = U_k^\dagger$~\cite{methods}, or equivalently, $\mathcal{K} H_{\text{eff}, k} \mathcal{K} = H_{\text{eff},k}$, where $H_{\text{eff},k}$ is the effective Hamiltonian satisfying $U_k \equiv e^{-iH_{\text{eff},k}T} $ with $T$ being the driving period. Without loss of generality, we set $T=1$. By spectral decomposition, we have $H_{\text{eff},k} = \sum_{n = 1,2,3}\epsilon_n |u_{n,k}\rangle\langle u_{n,k}|$ with $\epsilon_n\in(-\pi,\pi]$. The eigenvectors $|u_{n,k}\rangle$ are all real and can be represented as unit vectors on the unit sphere, and together form an orthogonal basis due to their mutual orthogonality. Changing the sign of each $|u_{n,k}\rangle$ leaves $H_{\text{eff},k}$ unchanged. Thus, the order-parameter space of the Hamiltonian is $M_3=O(3)/O(1)^3$, where $O(N)$ is the orthogonal group in $N$ dimensions. The first homotopy group of this space is $\pi_1(M_3) = Q_8$---a quaternion group with eight elements $\{\pm1,\pm I,\pm J, \pm K\}$, which is non-Abelian and satisfies $I^2=J^2=K^2=IJK=-1$, and the non-commutative relations $IJ=-JI$, $JK=-KJ$, and $KI=-IK$~\cite{wu,guo,lihu}.

In our experiment, we implement domain-wall configurations using quantum-walk dynamics. We therefore resort to the multifold bulk-boundary correspondence in the domain-wall setup (See Supplementary Text). In a non-Abelian system with the domain-wall boundary condition, we define the boundary states as the domain states. The domain-wall states are related to the domain-wall charge~\cite{guo,lihu}, which is defined as $\Delta Q = Q_L/Q_R$, and is the topological charge of the configuration. Here $Q_L$ and $Q_R$ are the charges of the left and right regions, respectively. In Fig.~\ref{fig1}B, we list the domain-wall charges in the first column, with the four classes of $Q_8$ in different rows. Possible spectral patterns of the domain-wall states corresponding to the domain-wall charges in the first column are illustrated in the second and third columns, in which the blue dots indicate the domain-wall states and the grey lines indicate bulk states.
Note that, in contrast to the static three-band model, which features only two band gaps, the Floquet system possesses an additional band gap crossing the first Brillouin zone boundary at $\pm\pi/T$, which significantly enriches the configurations of domain-wall states. We denote the bands from the bottom to top as the first, second, and third bands, with the gaps between them corresponding to the first, second, and third band gaps, respectively.

\begin{figure}
\includegraphics[width=0.5\textwidth]{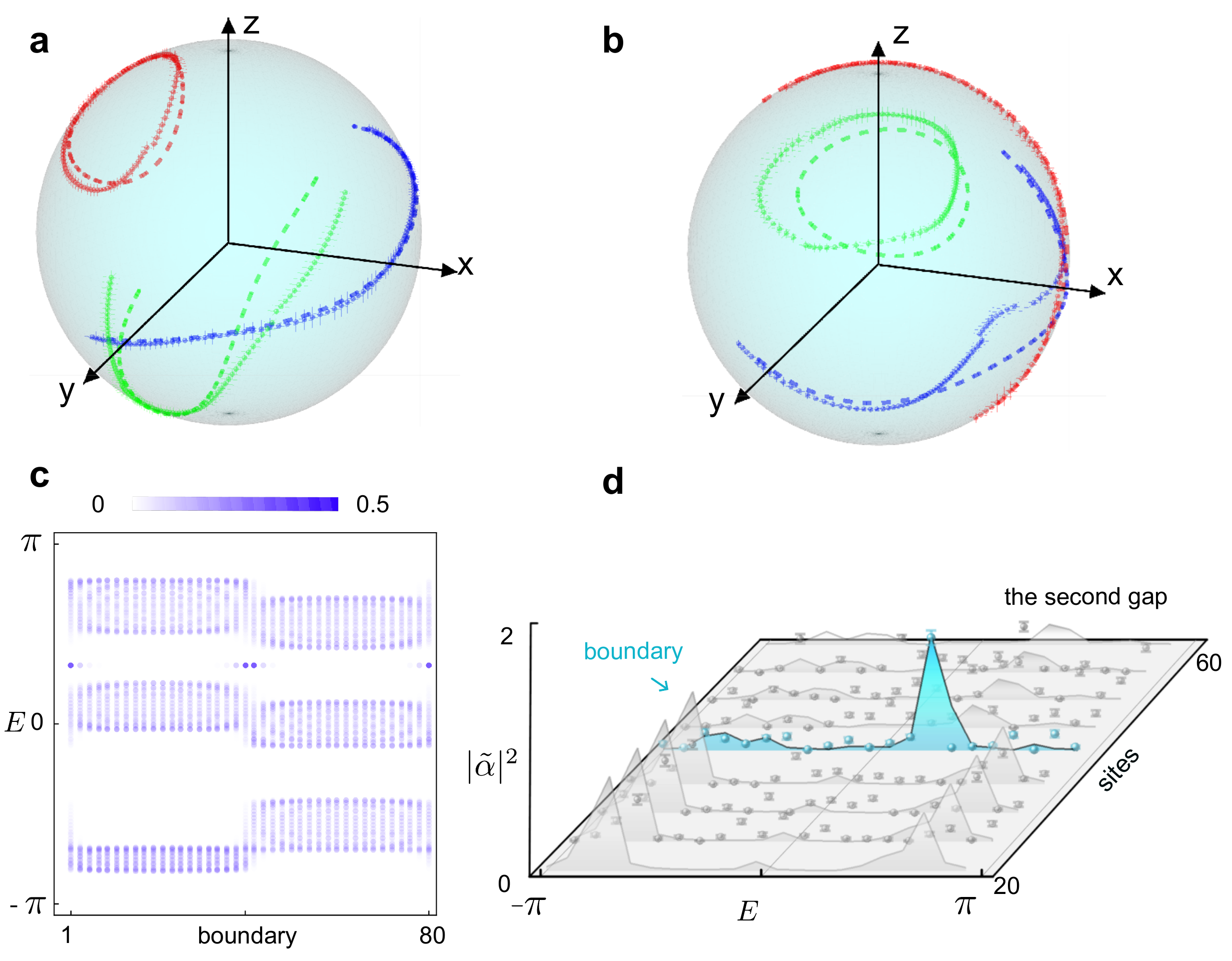}
\caption{{\bf Mutlifold bulk-boundary correspondence: domain-wall states within a single gap.} ({\bf A}), ({\bf B}) Trajectories of eigenstates of the left and right bulks, respectively, on the unit sphere by varying the momentum $k$ from $-\pi$ to $\pi$, the blue, green and red dashed lines/dots indicate the eigenstates of the first ($|u_{1,k}\rangle$), second ($|u_{2,k}\rangle$) and third ($|u_{3,k}\rangle$) bands, respectively. Dashed lines represent the theoretical predictions, while dots represent the experimental data. ({\bf A}) Each of the vectors $|u_{1,k=\pi}\rangle$ and $|u_{2,k=\pi}\rangle$ acquires an additional negative sign compared to the vectors at $k=-\pi$, indicating that the system has a quaternion topological charge $K$. ({\bf B}) Each of the vectors $|u_{1,k=\pi}\rangle$ and $|u_{3,k=\pi}\rangle$ acquires a negative sign compared to the vectors at $k=-\pi$, corresponding to quaternion topological charge $J$. ({\bf C}) The quasienergy spectra and spatial distributions of the corresponding eigenstates with domain-wall boundary conditions, the domain-wall states emerge within the second gap, corresponding to the domain-wall charge $I$ in Fig.~\ref{fig1}B (the second column). ({\bf D}) Observation of domain-wall state via the spatially-resolved injection spectroscopy method. Blue dots denote the initial state, which is injected at the domain wall. A prominent peak in $|\tilde\alpha|^2$ occurs at $E = E_b$, indicating the presence of the domain-wall state. Theoretical predictions are represented by the corresponding colored lines filling bottom. The parameters are listed in Table~\ref{tab1}. Error bars are due to the statistical uncertainty in photon-number-counting. }
\label{fig2}
\end{figure}

\begin{figure}
\includegraphics[width=0.5\textwidth]{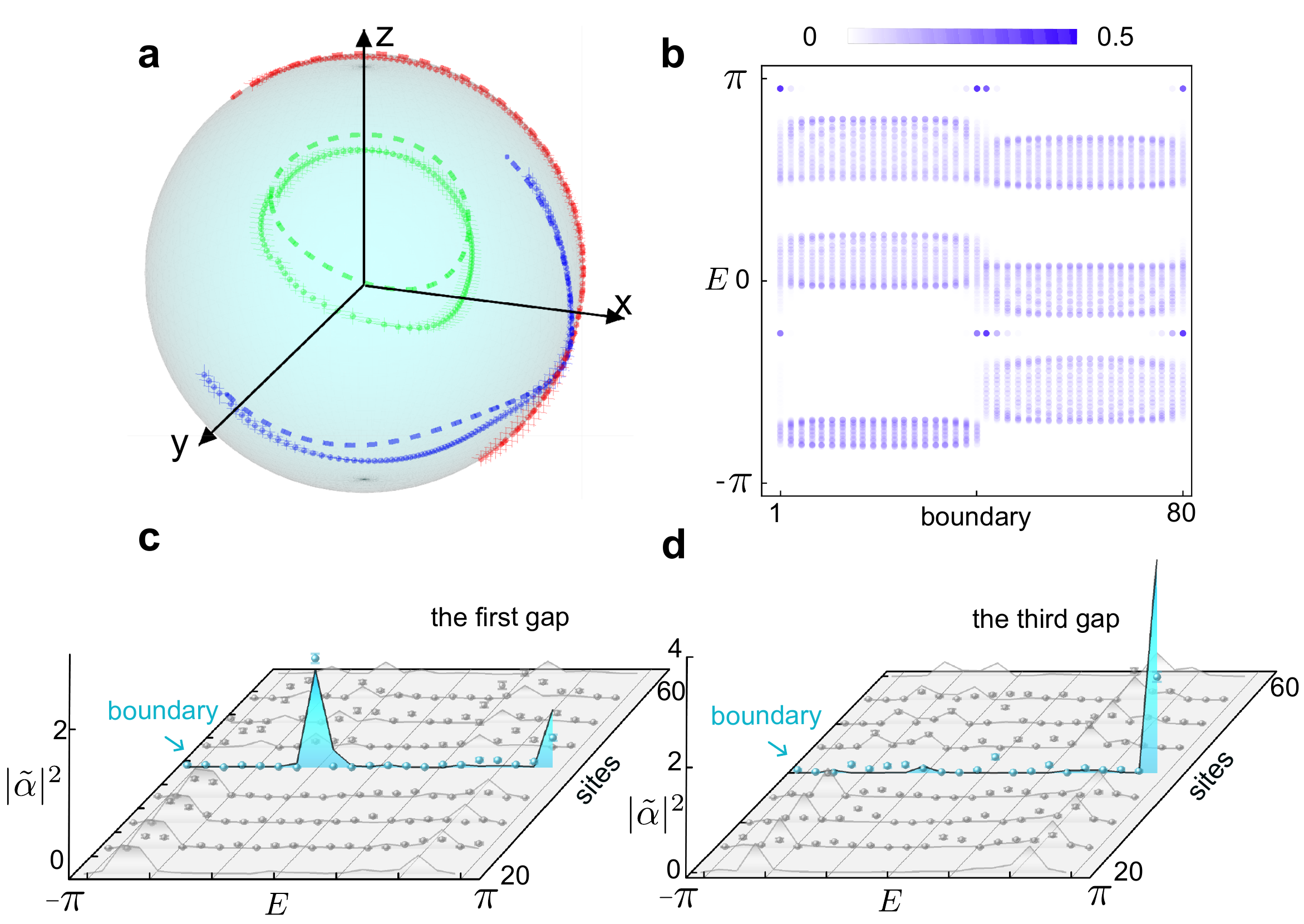}
\caption{{\bf Mutlifold bulk-boundary correspondence: domain-wall states within two gaps.} Parameters of the left bulk are the same as those in Fig.~\ref{fig2}A. ({\bf A}) Trajectories of eigenstates of the right bulk. Each of the vectors $|u_{1,k=\pi}\rangle$ and $|u_{3,k=\pi}\rangle$ acquires a negative sign compared to the vectors at $k=-\pi$, indicating that the system has a quaternion topological charge $J$. ({\bf B}) The domain-wall quasienergy spectra and spatial distributions of the corresponding eigenstates. The domain-wall states emerge within both the first and third gaps, corresponding to domain-wall charge $I$ in Fig.~\ref{fig1}B (the third column). ({\bf C}), ({\bf D}) Observation of the domain-wall states via the spatially-resolved injection spectroscopy method. The parameters are listed in Table~\ref{tab1}.
}
\label{fig3}
\end{figure}

\begin{figure}
\includegraphics[width=0.5\textwidth]{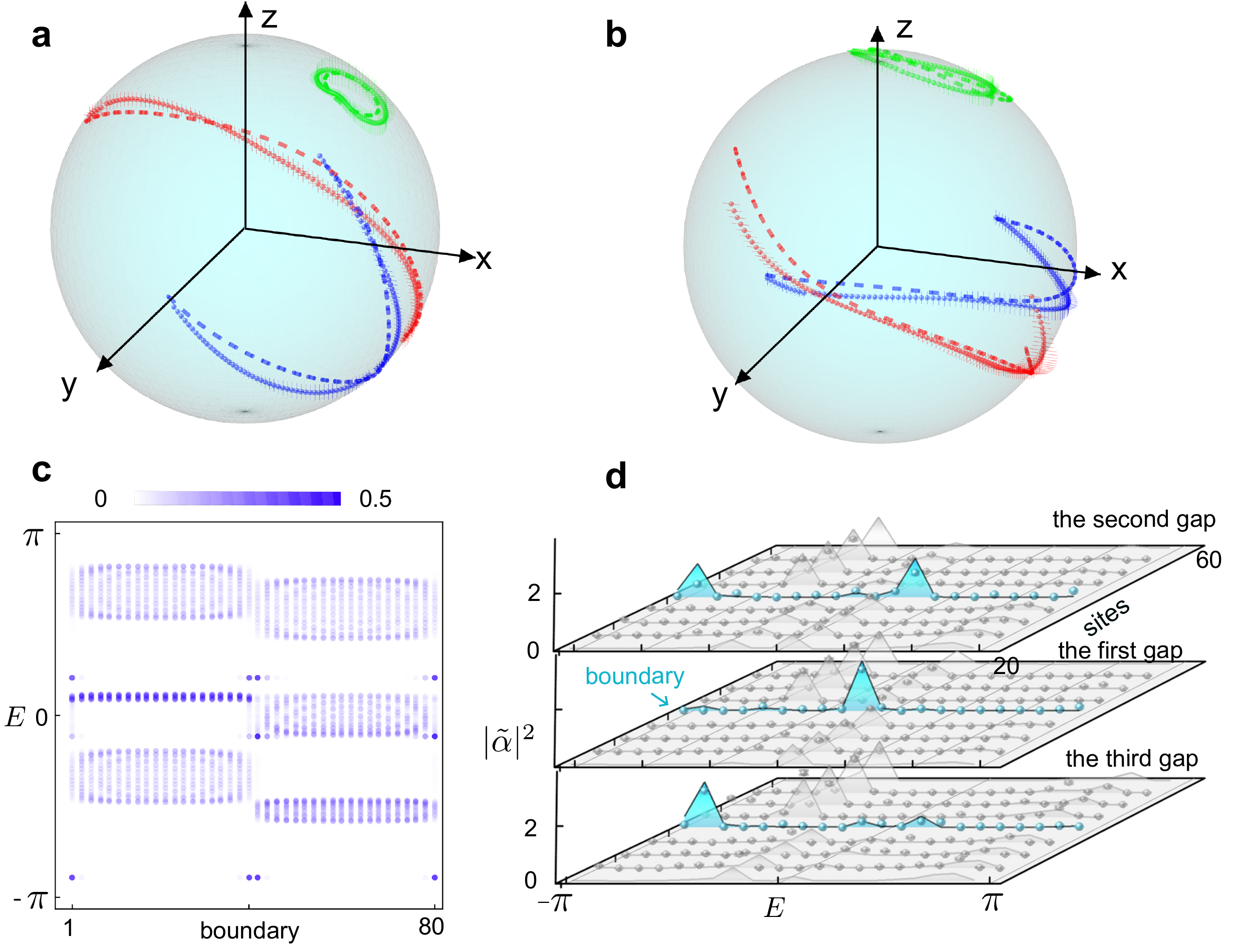}
\caption{{\bf Probing anomalous non-Abelian state.} ({\bf A}), ({\bf B}) Trajectories of eigenstates on the unit sphere by varying the momentum $k$ from $-\pi$ to $\pi$. In the left and right bulks, each of the vectors $|u_{1,k=\pi}\rangle$ and $|u_{3,k=\pi}\rangle$ acquires a negative sign compared to the vectors at $k=-\pi$, indicating that both sides of the bulk have quaternion topological charge $J$. This results in the domain-wall charge $1$. ({\bf C}) Quasienergy spectra and spatial distributions of the corresponding eigenstates with the domain-wall boundary condition. The domain-wall states emerge within all three band gaps. ({\bf D}) Observation of the domain-wall states via the spatially-resolved injection spectroscopy method. The parameters are listed in Table~\ref{tab1}.
}\label{fig4}
\end{figure}

{\bf Observation of multifold bulk-boundary correspondence.}
We first measure the quaternion topological charge of the system by constructing the time-evolution matrix $U_k$ from bulk dynamics. Initializing the walker in the state $|\psi(0)\rangle_A = |0\rangle \otimes |A\rangle$, we evolve it for one time step $U(t=T)$ and perform state tomography to access the wave function $|\psi(t=T)\rangle$. Then, by applying the Fourier transform, we obtain $U_k|A\rangle$~\cite{methods}.
We repeat the above process for different initial states $|\psi(0)\rangle_B = |0\rangle \otimes |B\rangle$ and $|\psi(0)\rangle_C = |0\rangle \otimes |C\rangle$, respectively, to arrive at the time-evolution matrix $U_k=(U_k|A\rangle, U_k|B\rangle, U_k|C\rangle)$.
By diagonalizing $U_k$, we obtain three real orthogonal vectors $\{|u_{n,k}\rangle\}_{n=1,2,3}$, corresponding to three trajectories on the unit sphere (see Figs.~\ref{fig2}A and B) when $k$ varies from $-\pi$ to $\pi$.
The topological structure of these trajectories encodes the quaternion topological charge.

In Figs.~\ref{fig2}A and B, we display the trajectories of the eigenstates on the unit sphere by varying $k$. These are marked in blue, green, and red for the first, second, and third bands, respectively. As shown in Fig.~\ref{fig2}A, each of the vectors $|u_{1,k=\pi}\rangle$ and $|u_{2,k=\pi}\rangle$ acquires an additional negative sign compared to the vectors at $k=-\pi$, indicating that the system has a quaternion topological charge $K$. In contrast, Fig.~\ref{fig2}B shows that each of the vectors $|u_{1,k=\pi}\rangle$ and $|u_{3,k=\pi}\rangle$ acquires a negative sign compared to the vectors at $k=-\pi$, indicating that the system has a quaternion topological charge $J$~\cite{methods}. We note that the sign of the charge inherently depends on the gauge of $|u_{n,k}\rangle$ $(n=1,2,3)$. For instance, for the model with charge $K$, if we change the signs of $|u_{1,k}\rangle$ and $|u_{3,k}\rangle$ while leaving $H_{\text{eff},k} = \sum_{n = 1,2,3}\epsilon_n |u_{n,k}\rangle\langle u_{n,k}|$ unchanged, the charge of the model changes to $-K$. Hence, in the following, we identify phases with charge $I$ ($J$ or $K$) and  $-I$ ($-J$ or $-K$) as the same.

We now focus on the domain-wall states and the multifold bulk-boundary correspondence. For this purpose, we connect two bulks with different quaternion topological charges in a ring configuration to enforce the domain-wall boundary condition. The domain-wall states emerge only within the specific energy gaps. As illustrated in Fig.~\ref{fig2}C, the domain-wall states are significantly localized at the boundaries, in contrast to the bulk states. To observe them, we develop a spatially-resolved injection spectroscopy, which yields the spectral information of the domain-wall states. This is crucial for fully characterizing our multigap system, where domain-wall states can appear within several gaps simultaneously.

We begin by preparing a localized initial state $|\psi(0)\rangle = \mathcal{N} \hat{P}_{\text{i,d}} |\psi_d\rangle$, where $|\psi_d\rangle$ represents the domain-wall state, $\hat{P}_{\text{i,d}} = |x_{i}\rangle \langle x_{d}|$ is the projection operator, where $x_d$ labels the domian-wall position and $x_i$ the position of injection. The factor $\mathcal{N}$ ensures the normalization of the initial state. Experimentally, we consider the scenario where only one boundary is accessible, and the other boundary is located far from the injection position. We then perform quantum-state tomography on the evolved state $|\psi(t)\rangle$, to extract the dynamical phase of the state for each time step $t$, which is directly related to the quasienergies of the domain-wall states~\cite{methods}. As the first step of the quantum-state tomography, we project the evolved state into the basis $|x_{i}\rangle \otimes |A\rangle$ ($|A\rangle$ being the initial coin state), and obtain $a(t) = (\langle A| \otimes \langle x_{i}|) |\psi(t)\rangle$. By varying $t$, we obtain $\alpha(t) = \{a(0), a(1), a(2), \ldots, a(t)\}$, where the maximum time step $t$ is $20$ in our experiment. The Fourier transform gives $\tilde{\alpha}(\omega)=\sum_{t'=0}^t \alpha(t') e^{i\omega t'}$, which maps $\alpha(t)$ from the time domain to the frequency domain, facilitating the identification of the dominant frequency (quasienergy) associated with $|\psi(t)\rangle$. When the injection is applied at the domain wall, a prominent peak in $|\tilde{\alpha}(\omega)|^{2}$ emerges near the quasienergies of the domain-wall states, as illustrated by the blue dots in Fig.~\ref{fig2}D. In contrast, when the initial state is injected away from the domain wall, the temporal evolution is governed by the bulk Hamiltonian. As a result, peaks of $|\tilde{\alpha}(\omega)|^{2}$ appear only at the quasienergies corresponding to the bulk states.

We first consider the case where the two bulks possess quaternion topological charges $K$ and $J$, respectively. Thus, the domain-wall charge is $\Delta Q = K/J = I$. The trajectories of the bulk eigenstates on the unit sphere are shown in Figs.~\ref{fig2}A and B. In Fig.~\ref{fig2}C, the domain-wall states emerge exclusively in the second gap, aligning with the configuration of the domain-wall charge $I$ in Fig.~\ref{fig1}B (the second column). This is confirmed by the experimental results of the injection spectroscopy (see Fig.~\ref{fig2}D), where a prominent peak emerges in the second gap.

By varying the parameters, we implement the domain-wall configuration with the domain-wall states located at the first and third band gaps, also accompanied by the domain-wall charge $I$ [see Fig.~\ref{fig1}B (the third column)]. In this case, we set the parameters of the left bulk to be the same as those in Fig.~\ref{fig2}A, and vary the parameters of the right bulk. The trajectories of the eigenstates traced on the unit sphere are shown in Fig.~\ref{fig3}A, indicating a quaternion topological charge $J$ in the right bulk. The quasienergy spectra and spatial locations of the corresponding eigenstates are shown in Fig.~\ref{fig3}B.
The domain-wall states emerge within both the first and third gaps, corresponding to the second configuration of domain-wall charge $I$ in Fig.~\ref{fig1}B.

In Figs.~\ref{fig3}C and D, we show the experimental results for different initial states via the injection spectroscopy, corresponding to the cases in which there is a significant overlap between the initial state and the domain-wall states within the first and third band gaps, respectively. Since the domain-wall states crossing multiple band gaps are mutually orthogonal, the spatially-resolved injection spectroscopy allows us to distinguish these states in different band gaps.
The measured peaks in Figs.~\ref{fig3}C and D are consistent with the quasienergy spectra in  Fig.~\ref{fig3}B. Their positions in the frequency domain match the quasienergies of the corresponding domain-wall states. We also provide experimental results for configurations with other domain-wall charges in Supplementary Text.

{\bf Anomalous non-Abelian phases.} Two-dimensional periodically driven systems can host topological edge modes even in the presence of trivial Chern numbers~\cite{rudner1,anoexp1,anoexp2,anoexp3}. A similar phenomenon occurs in one-dimensional Floquet non-Abelian systems, where topological edge modes exist in all band gaps despite
a trivial topological charge~\cite{lihu}. To demonstrate this anomalous behavior, we implement a domain-wall system, in which the quaternion topological charges of the left and right bulks are both $J$ (see Figs.~\ref{fig4}A and B), resulting in a trivial domain-wall charge $\Delta Q = J/J=1$. Based on the numerical calculations in Fig.~\ref{fig4}C, although the domain-wall charge is topological trivial, the domain-wall states emerge in all three band gaps. Experimentally, when the photon is injected into the boundary, the evolution is governed by the domain-wall states. The evolved state carries information of the domain-wall states' quasienergies, which can be obtained through the injection spectroscopy. As illustrated in Fig.~\ref{fig4}D, the peaks of $ |\tilde{\alpha}|^{2}$ emerge if and only if the domain-wall states exist, and are located near the quasienergies of the domain-wall states. By contrast, if the injection is far away from the domain wall, the peaks corresponding to the domain-wall states cannot be observed. Hence, we experimentally confirm the anomalous non-Abelian phases.

{\bf Discussion.} We experimentally implement the FNATI, and establish the multifold bulk-boundary correspondence. In particular, we identify the anomalous non-Abelian phases, in which topological edge states appear in all band gaps despite a trivial topological charge. Our experiment is facilitated by the implementation of quantum-walk dynamics with higher-dimensional coin states, as well as by the development of novel detection schemes such as the spatially-resolved injection spectroscopy. Our results highlight the versatility of quantum-walk dynamics as a tool for investigating non-Abelian topological phases, with potential applications in Floquet topological matter and beyond. Our experimental setup and detection schemes can also be extended to higher-band systems, offering access to intriguing possibilities, such as simulating multiband Floquet topological insulators characterized by the higher-dimensional non-Abelian group.

{\bf Methods}

{\bf $\mathcal{PT}$ symmetry of $U_k$.} The Floquet operator of the quantum walk in the momentum basis is
\begin{equation}
U_k = U_h RS_kR U_h,
\end{equation}
where $U_h = e^{-iM/2}$ with $M =
\begin{pmatrix}
a& r&u\\
r& h&s\\
u&s&g
\end{pmatrix}$ and $R=
\begin{pmatrix}
\cos(\theta)& i\sin(\theta)&0\\
i\sin(\theta)& \cos(\theta)&0\\
0&0&e^{i\phi}
\end{pmatrix}$ are two rotation operators,
and $ S_k=
\begin{pmatrix}
e^{ik}& 0&0\\
0& 1&0\\
0&0&e^{-ik}
\end{pmatrix}
$ is the position shift operator.

In an appropriate basis, the combination of the space-inversion $\mathcal{P}$ and time-reversal $\mathcal{T}$ symmetries can be represented by the complex conjugation operator $\mathcal{K}$.  A  $\mathcal{PT}$ symmetry Hamiltonian satisfies $H_k = \mathcal{K}H_k\mathcal{K} = H^*_k$. It can be shown that if two Hamiltonians $H_{1,k}$ and $H_{2,k}$ both possess the $\mathcal{PT}$ symmetry, then the effective Hamiltonian $H_{\text{eff}}$ satisfying $U_k = e^{-iH_{1,k}}e^{-iH_{2,k}}e^{-iH_{1,k}}\equiv e^{-iH_{\text{eff},k}}$ also possess the $\mathcal{PT}$ symmetry, as
\begin{align}
\mathcal{K} U_k \mathcal{K}&= e^{i\mathcal{K}H_{1,k}\mathcal{K}}e^{i\mathcal{K}H_{2,k}\mathcal{K}}e^{i\mathcal{K}H_{1,k}\mathcal{K}} \nonumber\\
&= e^{iH_{1,k}}e^{iH_{2,k}}e^{iH_{1,k}} = U^\dagger_k,
\end{align}
which implies $\mathcal{K} H_{\text{eff},k} \mathcal{K} = H_{\text{eff},k}.$

\begin{table}[!htbp]\centering
\begin{tabular*}{8cm}{@{\extracolsep{\fill}}|c|cccccccc|}
\hline
\quad $Q$ \quad & $r$ & $s$ & $u$ & $a$ & $h$ & $g$ & $\theta$ & $\phi$ \\
\hline
$K$~~(Fig.~\ref{fig2}{\bf{a}}) & $0$ & $1$ & $1$ & $0$ & $0$ & $0$ & $1$ & $0$ \\
\hline
$J$~~(Fig.~\ref{fig2}{\bf{b}}) & $0$ & $1$ & $1$ & $0$ & $0$ & $0$ & $0$ & $0$ \\
\hline
$J$~~(Fig.~\ref{fig3}{\bf{a}}) & $0$ & $1$ & $1$ & $0$ & $0$ & $0$ & $3$ & $0$ \\
\hline
$J$~~(Fig.~\ref{fig4}{\bf{a}}) & $2$ & $0$ & $2$ & $1$ & $0$ & $-1$ & $1.25$ & $-0.7$ \\
\hline
$J$~~(Fig.~\ref{fig4}{\bf{b}}) & $2$ & $0$ & $2$ & $1$ & $0$ & $-1$ & $-1.1$ & $-0.1$ \\
\hline
\end{tabular*}
\caption{Parameters used in the figures of the main text.}
\label{tab1}
\end{table}

{\bf Detect the quaternion topological charge via the bulk dynamics.} In the momentum basis, the time evolution operator can be written as
\begin{align}
U = \int_{-\pi}^{\pi}dk|k\rangle\langle k |\otimes U_k.
\end{align}
An arbitrary initial state in the momentum space is expressed as
\begin{align}
|\psi(0)\rangle = \sum_{k} C_{k} |k\rangle\otimes|v_{k}\rangle,
\end{align}
where $C_{k}$ is the coefficient of each $k$ component. For evolution over $t$ time steps, the state evolves as
\begin{align}
|\psi(t)\rangle = \sum_k C_k |k\rangle \otimes U_k^t|v_k\rangle.
\end{align}

If we consider an initially state
\begin{align}
|\psi(0)\rangle_A = |0\rangle\otimes|A\rangle,
\end{align}
which is localized and can be written in the momentum space as $|\psi(0)\rangle_A = \frac{1}{\sqrt{N}}\sum_{k} |k\rangle\otimes|A\rangle$. After one step of the quantum walk, the state evolves to $|\psi(1)\rangle_A = \frac{1}{\sqrt{N}}\sum_{k} |k\rangle\otimes U_k|A\rangle$. If the state is projected onto the $k$-space, for a fixed $k$, we obtain $U_k|A\rangle$. Similarly, if we consider different initial states, $|\psi(0)\rangle_B = |0\rangle\otimes|B\rangle$ and $|\psi(0)\rangle_C =  |0\rangle\otimes|C\rangle$, with the same process, we obtain $U_k|B\rangle$ and $U_k|C\rangle$, respectively. By collecting these results, we construct the matrix
\begin{align}
U_k=(U_k|A\rangle, U_k|B\rangle, U_k|C\rangle).\label{eq:Uk}
\end{align}

By diagonalizing $U_k$, we obtain three real orthogonal vectors $\{|u_{n,k}\rangle\}$ ($n=1,2,3$). As $k$ varies from $-\pi$ to $\pi$, these vectors trace three trajectories on the unit sphere. The topological structure of these trajectories encodes the quaternion topological charge in Ref~\cite{lihu}.

When $k$ varies from $-\pi$ to $\pi$, i) if the vectors $|u_{2,k}\rangle$ and $|u_{3,k}\rangle$ rotate from $\pi$ to $-|u_{2,k}\rangle$ and $-|u_{3,k}\rangle$, respectively, then the system has a quaternion topological charge $I$; ii) if $|u_{1,k}\rangle$ and $|u_{3,k}\rangle$ rotate from $\pi$ to  $-|u_{1,k}\rangle$ and $-|u_{3,k}\rangle$, respectively, then the quaternion topological charge is $J$; iii) if $|u_{1,k}\rangle$ and $|u_{2,k}\rangle$ rotate from $\pi$ to  $-|u_{1,k}\rangle$ and $-|u_{2,k}\rangle$, respectively, the quaternion topological charge is $K$; iv) if two of the three eigenvectors $|u_{1,k}\rangle, |u_{2,k}\rangle$ and $|u_{3,k}\rangle$ rotate $2\pi$ while all three vectors return to themselves, the quaternion topological charge of this system is $-1$. Thus, the quaternion topological charges are encoded in the topological structure of eigenstate trajectories.

On the other hand, the rotations of the eigenstate with the momentum $k$ varying from $-\pi$ to $\pi$ can be characterized by the quaternion topological charge $Q\in Q_8$, analytically. We introduce the Wilson operator and define the Wilson loop along the Brillouin zone $B$ as
\begin{align}
W_B = \mathbb Pe^{ \oint_B A_{\text{all},k} dk},
\end{align}
where $[A_{\text{all},k}]_{mn}=\langle u_{m,k}|\partial_k|u_{n,k}\rangle$ represents the Berry-Wilczek-Zee connection, $m$ and $n$ are the band indices~\cite{wu}, and $ \mathbb P$ is the path-ordering operator. Numerically, the operator $ \mathbb P$ satisfies $ \mathbb Pe^{ \oint_B A_{\text{all},k} dk} = \mathbb P \{\Pi_{\{k_\alpha \}_{\alpha=1}^\Lambda}e^{ A_{\text{all},k_\alpha} \Delta k}\} = e^{ A_{\text{all},k_1} \Delta k}e^{ A_{\text{all},k_2} \Delta k} \cdots e^{ A_{\text{all},k_\Lambda}\Delta k}$, where $ \{k_\alpha \}_{\alpha=1}^\Lambda$ corresponds to a sequence of $k$-points ordered along the Brillouin zone $B$, $\Delta k$ is the distance between two sequential $k$-points, and $\Lambda $ is the total number of $k$-points on the Brillouin zone.
The operator $A_{\text{all},k}$ is anti-symmetric and can be decomposed into the ${\mathfrak {so}}(3)$ Lie-algebra basis
\begin{align}
A_{\text{all},k}=\sum_{i=1,2,3}\beta_iL_i,
\end{align}
where $(L_i)_{jk}=-\epsilon_{ijk}$, and $\epsilon_{ijk}$ denotes the anti-symmetric tensor. Lift $A_\text{all}$ to the $\mathfrak{spin}(3)$-valued 1-form by replacing $L_i$ with $t_i$, where $t_i=-\frac{i}{2}\sigma_i$ and $\sigma_i$ represents the Pauli matrix, we have
\begin{align}
\overline A_{\text{all},k}=\sum_{i=1,2,3}\beta_it_i.
\end{align}
Then, the non-Abelian charge is defined as
\begin{align}
Q=\mathbb Pe^{ \oint_\Gamma \overline A_{\text{all},k} dk},
\end{align}
which is a $2\times 2$ matrix with the form $\pm i\sigma_n$ ($n = 0,1,2,3$) and $\sigma$ are the Pauli matrices. The elements of the quaternion group are represented as follows $\pm I\rightarrow \mp i\sigma_x$, $\pm J\rightarrow\mp i\sigma_y$, $\pm K\rightarrow\mp i\sigma_z$ and $\pm1\rightarrow \mp i \sigma_0$. In the experiment, we reconstruct $U_k$ in Eq.~(\ref{eq:Uk}) via quantum-state tomography. The eigenstate trajectories and their topological structures can be calculated through $U_k$, leading to the detection of the charge.

{\bf Experimental realization.} We implement the photonic quantum walk using the time-multiplexed approach. We extend the well-studied two-band quantum-walk model~\cite{lq1,lq2,sch} to a three-band quantum-walk model. This extension is important for investigating more complex topological phenomena, particularly non-Abelian bulk-boundary correspondence. The core of our experimental setup involves the construction of the rotation operator $U_h$, belonging to the $U(3)$ group. To achieve this, we decompose $U_h$ into the product of three $U(2)$ matrices, which manipulate in the two-dimensional subspaces and act as an identity operator on the remaining dimension. The decomposition is given by
\begin{align}
U_h = U_{23} U_{13} U_{12}.
\end{align}
In our experiment, two triple PBSs together with waveplates are used to merge the photons in two different spatial modes into one spatial mode. Then we apply waveplates on the photons in the spatial mode to manipulate the polarizations of the photons. Thus, we realize the $SU(2)$ transformation in arbitrary two-dimensional subspaces~\cite{zx}. As the $U(2)$ transformation differs from $SU(2)$ by a global phase, which can be adjusted by the optical path differences between different arms of our uniquely constructed interferometer, thus we can realize $U_{ij}$ with our setup.

We rewrite the evolution operator of quantum walks as $U^{t} = U_h U_{\text{loop}}^{t} U_h^{-1}$ with $U_{\text{loop}} = R S R U_h^2$, which allows us to implement the dynamics within a fiber loop configuration. The rotation operator $R$ is realized by two electro-optic modulators (EOMs) with different optical axis angles. The birefringent crystal inside each EOM is set at a $45^\circ$ angle to the $x/y$ axis (aligned with the laboratory coordinates). So that the EOM acts on the polarizations of photons as follows
$R_\text{EOM}(\theta) =
\begin{pmatrix}
1 & 1 \\
-1 & 1
\end{pmatrix}
\begin{pmatrix}
e^{i\theta} & 0 \\
0 & e^{-i\theta}
\end{pmatrix}
\begin{pmatrix}
1 & -1 \\
1 & 1
\end{pmatrix}
=
\begin{pmatrix}
\cos\theta & i\sin \theta \\
i\sin\theta & \cos\theta
\end{pmatrix}.$
We place the EOM in the upper spatial mode and another EOM with the optical axis set to $0^\circ$ in the lower spatial mode to implement the phase modulation $e^{i\phi}$ for the state $|DH\rangle$. The shift operator $S$ is implemented by separating photons into different basis states using a set of PBSs and routing them through fibers of different lengths to introduce a well-defined time delay. Specifically, photons in the states $|UH\rangle$, $|UV\rangle$ and $|DH\rangle$  pass through Paths 1, 2 and 3, respectively. The corresponding time delays are $1,468.13$ns, $1,551.53$ns, and $1,634.93$ns, respectively. The resulting temporal differences correspond to spatial modes $x+1, x, x-1$, respectively. Finally, by applying $U_h^{-1}$ at the initial-state preparation stage and $U_h$ before the detection, we experimentally implement the Floquet operator $U$.

{\bf Quantum-state tomography.} For the detection of the multifold bulk-boundary correspondence, we reconstruct the evolved state $|\psi(t)\rangle = U^{t}|\psi(0)\rangle$ at each time step, as all observables mentioned in the main text are based on $|\psi(t)\rangle$. Considering that all selected initial states in the main text are pure states, and given the high stability of the experimental setup, the evolved states can be assumed as pure states too. The evolved state can be expressed as~\cite{xiaoy}
\begin{align}
|\psi(t)\rangle &= e^{i\phi'_t} \sum_{x} \Big[ P_{UH}(t,x) |x\rangle \otimes |UH\rangle  \\
&+ P_{UV}(t,x) |x\rangle \otimes |UV\rangle + P_{DH}(t,x) |x\rangle \otimes |DH\rangle \Big],\nonumber
\end{align}
where $e^{i\phi'_{t}}$ represents the relative phase for different time steps $t$. We then implement four distinct measurement protocols to reconstruct $|\psi(t)\rangle$ in the basis $\left\{ |UH\rangle, |UV\rangle, |DH\rangle, |DV\rangle \right\}$.

Protocol-{\bf{1}}. We measure the absolute values $\left|p_\mu(t,x)\right|$ $(\mu=UH,UV,DH)$. After the $t$-th time step, photons in the position $x$ arrive at the first detection device, $M_1$, which consists of PBSs and APDs. This device performs a projective measurement in the basis {$|UH\rangle, |UV\rangle, |DH\rangle$}. The measured probability distributions are given by:
\begin{align}
|P_\mu(t,x)|^{2}=\frac{ N_\mu(t,x)}{\sum_{x}\left[N_{UH}(t,x)+N_{UV} (t,x)+ N_{DH} (t,x)\right]},
\end{align}
where $N_\mu(t,x)$ represents the photon counts.

Protocol-{\bf{2}}: Without loss of generality, we rewrite the evolved state as $|\psi(t)\rangle=\sum_{x}|\psi_{x}(t)\rangle \otimes |x\rangle$, where $|\psi_{x}(t)\rangle = e^{i\tilde{\phi}_{x}}(|P_{UH}(t,x)| |UH\rangle + e^{i\phi_{x,UV}}|P_{UV}(t,x)| |UV\rangle + e^{i\phi_{x,DH}}|P_{DH}(t,x)| |DH\rangle)$. The goal here is to measure the phases $\phi_{x,UV}$ and $\phi_{x,DH}$. Photons at the position $x$ are directed to the second detection device, $M_2$, which consists of two EOMs and the first detection device $M_1$. The first EOM acts on the polarization of photons as $\begin{pmatrix}
e^{i\theta_1} & 0 \\
0 & e^{-i\theta_1}
\end{pmatrix}$ and the second EOM acts as $\begin{pmatrix}
\cos\theta_2 & -\sin \theta_2 \\
\sin\theta_2 & \cos\theta_2
\end{pmatrix}$. These operations can be achieved by adjusting the appropriate optical axis angles of the EOMs and the waveplates~\cite{lq2}. We set the voltage of the second EOM such that $\tan(\theta_2)=|P_{UH}|/|P_{UV}|$, using the data obtained in protocol-{\bf{1}}. To measure $e^{i\phi_{x,UV}}$, we sweep $\theta_1$ by controlling the voltage of the first EOM and observe the interference between the photons in the states $|UH\rangle$ and $|UV\rangle$. We have $\theta_{1m}=\phi_{x,UV}/2$, when the minimum photon count in the state $|UH\rangle$ is observed~\cite{bongk}. Thus, we can determine $e^{i\phi_{x,UV}}$ by reading the voltage of the first EOM. Similarly, we can measure $e^{i\phi_{x,DH}}$. We need to combine the two modes $|UV\rangle, |DH\rangle$ into one spatial mode. This is achieved by a triple PBS and waveplates in front of the measurement device. Then, we obtain the phase information $\phi_{x,UV}$ and $\phi_{x,DH}$ from Protocol-{\bf{2}}.

Protocol-{\bf{3}}. We probe the relative phase between the amplitudes $P_{UH}(t,x)$ and $P_{UH}(t,x')$.  For example, we consider the relative phase between the amplitudes at positions $x$ and $x+1$. First, we introduce a detection unit $M_3$, which consists of an additional loop and the detection unit $M_2$. In the extra loop, we turn off the EOMs in the fiber loop, so that the operation $R$ is replaced by an identity operation. We also remove the operator $U_h$ before the detection. Then we have
\begin{align}
\left\langle UH \mid \psi_x(t+1)\right\rangle &=e^{i \tilde{\phi}_x} e^{i \Delta\tilde{\phi}_x}(\tilde{a} |P_{UH}(t, x+1)|\\
&+ \tilde{b} e^{i \phi_{x+1,UV}}|P_{UV}(t, x+1)| \nonumber\\
&+ \tilde{c} e^{i \phi_{x+1,DH}}|P_{DH}(t, x+1)|), \nonumber\\
\left\langle UV \mid \psi_x(t+1)\right\rangle &= e^{i \tilde{\phi}_x} (\tilde{d} |P_{UH}(t, x)| \nonumber\\
&+ \tilde{e} e^{i \phi_{x,UV}} |P_{UV}(t, x)| \nonumber\\
&+ \tilde{f} e^{i \phi_{x,DH}}|P_{DH}(t, x)|),\nonumber
\end{align}
where $e^{-i M / 2}=\left(\begin{array}{ccc}\tilde{a} & \tilde{b} & \tilde{c} \\ \tilde{d} & \tilde{e} & \tilde{f} \\ \tilde{g} & \tilde{h} & \tilde{i}\end{array}\right)$ and $e^{i \Delta\tilde{\phi}_x}=e^{i (\tilde{\phi}_{x+1}-\tilde{\phi}_{x})}$. Since the phases of $e^{i \phi_{x,UV}}$ and $e^{i \phi_{x,DH}}$ have been obtained in Protocol-{\bf{2}}, we can extract $e^{i \Delta\tilde{\phi}_x}$ using the same method in Protocol-{\bf{2}}.

Protocol-{\bf{4}}. To detect the relative phase for different time steps $t$, we extend the coin state to four dimensions by introducing a reference dimension $|DV\rangle$. We perform identity operator to photons in the state $|DV\rangle$ and guide them always to the right spatial mode. For every initial state prepared in the main text, we extend it to $\tilde{\psi}(0)=\mathcal{N}\begin{pmatrix}\psi(0) \\1\end{pmatrix}$, where $\mathcal{N}$ is a normalization factor. The state $\tilde{\psi}(0)$ evolves under
$\tilde{U}_k=\begin{pmatrix}
 U_k & 0\\
 0 & 1
\end{pmatrix}$. After each step, we measure the relative phase between the amplitudes $P_{UH}(t,x=t)$ and $P_{DV}(t,x=t)$ via the same method in Protocol-{\bf{2}}, which allows us to extract $e^{i\tilde{\phi}_{x=t}}$. By combining this with the recurrence relation obtained in Protocol-{\bf{3}}, we can determine the phase information $e^{i\tilde{\phi}_{x}}$ for arbitrary $x$. Finally, $|\psi(t)\rangle$ can be reconstructed by four distinct measurement protocols.

{\bf Spatially-resolved injection spectroscopy method.} To investigate non-Abelian bulk-boundary correspondence, the domain-wall boundary condition is enforced by modulating the EOMs to vary the parameters $\theta$ and $\phi$ of the quantum-walk system. Then, we choose an initial state with significant overlap a domain-wall state, $|\psi(0)\rangle = \mathcal{N} \hat{P}_{\text{d}} |\psi_d\rangle$, where $\hat{P}_{\text{d}} = |x_{d}\rangle \langle x_{d}|$ is the projection operator, $x_d$ is the domain-wall position, and $|\psi_d\rangle$ represents the domain-wall state. Since the domain-wall state is primarily localized at $x_d$, its overlap with the initial state is approximately $|\langle \psi_d | \psi(0) \rangle| \approx 1$, while the overlap between the initial state and any extended bulk state $|\psi_n\rangle$ is negligible, i.e., $|\langle \psi_n | \psi(0) \rangle| \approx 0$. Consequently, after $t$ steps, the temporal evolution of the state is $|\psi(t)\rangle = U^t |\psi(0)\rangle = \sum_{m}  e^{-iE_mt}|\psi_m\rangle\langle \psi_m |\psi(0)\rangle \approx e^{-iE_dt} |\psi_d\rangle$, where $E_d$ represents the quasienergy of the domain-wall state. To extract the dynamical phase $e^{-iE_dt}$, we perform quantum-state tomography on the evolved state $|\psi(t)\rangle$ at each time step. This results in a series of phases $\{ e^{-iE_dt'} \}_{t' = 0,1,\cdots,t}$, and their Fourier transformation is $\tilde{\alpha}(\omega) = \sum_{t'=0}^{t} e^{-iE_dt'} e^{i\omega t'}$. The amplitude of $\tilde{\alpha}(\omega)$ reach its maximum value at $\omega = E_d$, which allows us to identify the quasienergy of the domain-wall state.

In our experiment, we choose different initial states which are injected at various positions $x_i$, including the bulk and domain-wall locations, that is, $|\psi(0)\rangle = \mathcal{N} \hat{P}_{\text{i,d}} |\psi_d\rangle$ with $\hat{P}_{\text{i,d}} = |x_{i}\rangle \langle x_{d}|$. We project the evolved state $|\psi(t)\rangle$ onto the state $|x_{i}\rangle|A\rangle$ and obtain $a(t) = \langle A|\langle x_i |\psi(t)\rangle$. The choice of the internal coin state only alters the amplitude of the projection, but does not affect the phase information. Therefore, for simplicity, we fix the internal coin state $|A\rangle$. By varying $t$, we obtain a set of projections, $\alpha(t) = \{a(0), a(1), a(2), \ldots, a(t)\}$. The Fourier transformation gives $\tilde{\alpha}(\omega)=\sum_{t'=0}^t \alpha(t') e^{i\omega t'}$. When the injection occurs near the domain wall, the prominent peaks of $|\tilde{\alpha}(\omega)|^2$ correspond to the quasienergies of the domain-wall states. In contrast, when the initial state is injected away from the boundary, the temporal evolution is governed by the bulk Hamiltonian. We can only observe peaks corresponding to the quasienergies of the bulk states, as the localized initial state overlaps only with the bulk states.

{\bf Data availability}

The data that support the findings of this study are available from the corresponding authors upon requests.

{\bf Code availability}

The codes that support the findings of this study are available from the corresponding authors upon requests.

{\bf Acknowledgments}

This work has been supported by the National Key R\&D Program of China (Grant Nos. 2023YFA1406701 and 2023YFA1406704) and National Natural Science Foundation of China (Grant Nos. 92265209, 12025401, 12374479). H.H. is supported by National Key Research and Development Program of China No. 2022YFA1405800 and National Natural Science Foundation of China (Grant No. 12474496).

{\bf Author contributions}

Q.L. performed the experiments and wrote part of the paper. T.L. and H.H. developed the theoretical aspects and wrote part of the paper. W.Y. performed the theoretical analysis, and wrote part of the paper. P.X. supervised the project, designed the experiments, analyzed the results and revised the paper.

{\bf Competing interest declaration}

The authors declare no competing interests.

{\bf Additional information}

Correspondence and requests for materials should be addressed to Wei Yi (wyiz@ustc.edu.cn) and Peng Xue (gnep.eux@gmail.com).

\end{document}